\documentclass[aps,twocolumn,showpacs,amsmath,amssymb,floatfix]{revtex4}
\usepackage{amsthm,times}
\usepackage{latexsym}
\usepackage{amsfonts}
\usepackage{bbm,dsfont}
\usepackage{graphicx}
\newcommand{\R}{\mathbb{R}}
\newcommand{\hil}{\mathcal{H}}
\newcommand{\tr}[1]{\mathrm{Tr}\left[ {#1} \right]} 
\newcommand{\Tr}[2]{\mathrm{Tr}_{#1}\left[ {#2} \right]} 
\newcommand{\ket}[1]{\left\vert {#1} \right\rangle} 
\newcommand{\bra}[1]{\left\langle {#1} \right\vert} 
 
\newcommand{\ketbra}[2]{\left\vert {#1} \right\rangle\left\langle {#2} \right\vert}  
\newcommand{\JC}{{\hbox{\small\sc jc}}}
\newcommand{\p}{{\scriptscriptstyle P}}
\newcommand{\Q}{{\scriptscriptstyle Q}}
\newcommand{\F}{{\scriptscriptstyle F}}
\begin{document}
\title{Qubit thermometry for micromechanical resonators}
\author{Matteo Brunelli}
\email{matteo.brunelli@studenti.unimi.it}
\affiliation{Dipartimento di Fisica, Universit\`a degli Studi di Milano, 
I-20133 Milano, Italy}
\author{Stefano Olivares}
\email{stefano.olivares@ts.infn.it}
\affiliation{Dipartimento di Fisica, Universit\`a degli Studi di
Trieste, I-34151 Trieste, Italy}
\affiliation{CNISM, UdR Milano, I-20133 Milano, Italy}
\author{Matteo G. A. Paris}
\email{matteo.paris@fisica.unimi.it}
\affiliation{Dipartimento di Fisica, Universit\`a degli Studi di Milano, 
I-20133 Milano, Italy}
\affiliation{CNISM, UdR Milano, I-20133 Milano, Italy}
\date{\today}
\begin{abstract}
We address estimation of temperature for a micromechanical oscillator 
lying arbitrarily close to its quantum ground state. Motivated by 
recent experiments, we assume that the oscillator is
coupled to a probe qubit via Jaynes-Cummings interaction and that
the estimation of its effective temperature is achieved via quantum
limited measurements on the qubit.  We first consider the ideal
unitary evolution in a noiseless environment and then take into
account the noise due to non dissipative decoherence.  We exploit 
local quantum estimation theory to assess and optimize the precision of
estimation procedures based on the measurement of qubit population, and to
compare their performances with the ultimate limit posed by quantum
mechanics.  In particular, we evaluate the Fisher information (FI)
for population measurement, maximize its value over the possible
qubit preparations and interaction times, and compare its behavior
with that of the quantum Fisher information (QFI). We found that the
FI for population measurement is equal to the QFI, i.e., population
measurement is optimal, for a suitable initial preparation of the
qubit and a predictable interaction time. The same configuration also 
corresponds to the maximum of the QFI itself.  
Our results indicate that the achievement of the ultimate bound to precision
allowed by quantum mechanics is in the capabilities of the current
technology. 
\end{abstract} 
\maketitle
\section{Introduction}
The edge between classical and quantum description of a phenomenon is
related to the interactions occurring between the system under
investigation and its environment. As a consequence, if we could, in
ideal conditions, avoid irreversible interactions among them we should
observe the emergence of quantum behavior even in macroscopic systems.
As a matter of fact, the technological developments of the recent years
have made it possible to start inquiring into the quantum limit even in
mesoscopic mechanical systems and experiments have been designed which
realize a solid state analogue of cavity quantum electrodynamics.  Many
of these experiments focus on detecting the quantization of vibrational
modes in a mechanical oscillator
\cite{nmr1,nmr2,nmr3,nmr4,nmr5,nmr6,nmr7,nmr8,nmr9,nmr10,nmr11}.
Experimental conditions such that a mechanical object may behave in a
quantum fashion are achieved in the low temperature regime. For example,
for a single vibrational mode of energy $\hbar \omega$  to show quantum
features, as the quantization of lattice vibrations, temperatures $T \ll
\frac{\hbar \omega}{k_B}$ are required, which for a micro-sized object
oscillating in the microwave band correspond to few mK. 
\par
In this framework it has become increasingly relevant to have a precise
determination of the temperature. However, for a quantum system in
equilibrium with a thermal bath, there is no linear operator that acts
as an observable for temperature. Temperature, thought as a macroscopic
manifestation of random energy exchanges between particles, still
retains its meaning but we have lost any operational definition. This
kind of impediment often occurs in physics, and especially in quantum
mechanics, whenever one is interested in quantities which are not directly
accessible, i.e. they do not correspond to observable quantities.  This
may either be due to experimental impossibilities, or be a matter of
principle, as it happens for nonlinear functions of the density
operator. In both cases, it turns out that the only way to gain some
knowledge about the quantity of interest is to measure one or more
proper observables somehow related to the parameter we are interested
in, and upon suitably processing the outcomes, to come back and infer
its value. Hence, any conceivable strategy aimed to evaluate the
quantity of interest ultimately reduces to a parameter estimation
problem. Relevant examples of this situation are given by estimation of
the quantum phase of a harmonic oscillator \cite{Mon06,HDB09,asp09,PD11}, the
amount of entanglement of a bipartite quantum state
\cite{EE08,EE10,EEL11} and the coupling constants of different kinds of
interactions \cite{Sar06,Hot06,Mon07,Fuj01,Zhe06,Boi08,ZP07,Cam10,Pat06,
mon10,mon11}. Here we
focus on the estimation of temperature \cite{Man89} and, motivated by recent
experimental achievements \cite{nmr11}, we specifically refer to schemes
where a micromechanical resonator is coupled to a superconducting
qubit, and then a measurement of the excited state  population
is performed on the qubit itself. From the statistics of the
population measurement is then possible to obtain information about
the oscillator state, e.g. infer how close it is to the ground
state, and in turn its temperature.
\par
In this context an optimization problem naturally arises, aimed at 
finding the most efficient inference procedure leading to minimum
fluctuations in the temperature estimate. In this paper we address 
this problem in the framework of local quantum estimation theory 
(QET) \cite{lqe1,lqe2,lqe3,lqe4,lqe5,lqe6}.  
We solve the dynamics of the qubit-resonator coupled system and, in order
to match realistic scenarios, we also take into account an effective
model for non dissipative decoherence. Then, we evaluate the Fisher
information (FI) for the estimation of temperature via population 
measurement (hereafter referred to as the FI {\em of the population
measurement}) and find both the
optimal initial qubit preparation and the smallest temperature value
that can be discriminated. Moreover, we evaluate the Quantum Fisher
Information (QFI) in terms of the symmetric logarithmic derivative in order to
calculate the ultimate bound to precision allowed by quantum mechanics.
This enable us to show that population measurement is
indeed optimal for a suitable choice of the initial preparation of the
qubit, and to provide quantum benchmarks for temperature
estimation.  
\par
It is worth noting at this point that we are not discussing here
temperature fluctuations in a thermodynamical setting. Although
temperature itself may not fluctuate, as it is suggested by quantum
thermodynamical approaches \cite{qth}, we expect that fluctuations
always appear in the temperature estimates coming from indirect measurements
\cite{web07,Jan11}. Quantum estimation theory provides the tools to evaluate lower bounds to
the amount of fluctuations for a given measurement, as well as the
ultimate bounds imposed by quantum mechanics.
\par
The paper is structured as follows. In Sec.  \ref{s:model} we describe
the interaction model:  first we briefly review the unitary Jaynes-Cummings 
dynamics for the coupled system and describe the measurements performed 
on the qubit, and then we take into account the decoherence effects. In
Sec. \ref{s:QET} we show how QET techniques applies to our system, 
providing explicit formulas for both the FI and the QFI. The results 
are finally shown in detail in Sec. \ref{s:results} both for the unitary
and the noisy dynamics. Sec. \ref{s:out} closes the paper with some
concluding remarks.   
\section{The physical model}\label{s:model}
As the temperature decreases a mechanical oscillator starts to exhibit 
its quantum nature, which mainly manifests itself in quantization of the 
vibrational modes. Hence, for our purposes the resonator can be regarded  
as a collection of phonons in a thermal equilibrium state. 
We assume that the resonator is built as to display an isolated 
mechanical mode at a given frequency, so that it can be modeled, rather 
than a phonon bath with some spectral distribution, as a single mode 
phonon field in thermal equilibrium.
\subsection{Unitary dynamics}
Let $\hil_R$ be the infinite dimensional Hilbert space
associated with the single mode phonon field. Upon introducing
the creation and annihilation operators $[a,a^{\dagger}]=1$ 
one has the number operator $N=a^{\dagger}a$,
and its eigenstates $\left\{
\ket{n}\right\}_{n=0}^{\infty}$. The field Hamiltonian reads:
\begin{equation}
H_\F = \hbar\, \Omega\, a^{\dagger}a \; ,
\end{equation}
where $\Omega$ denotes the frequency of the vibrational mode. 
We assume the resonator in a thermal equilibrium state, i.e. 
described by the density operator
\begin{eqnarray}\label{thermal}
\varrho_{F}&=&\frac{\exp(-\beta H_\F)}{\tr{\exp(-\beta H_\F)}} 
= \sum_{n=0}^{\infty} p_n (\Omega, \beta)
\ketbra{n}{n} \; , \nonumber
\end{eqnarray}                   
where $\beta=(k_B T)^{-1}$ and:
\begin{equation}\label{pn}
p_n (\Omega, \beta) =
e^{-\beta \hbar \Omega n}\left(1- e^{-\beta \hbar \Omega}\right).
\end{equation}                  
The resonator is coupled to a superconducting qubit whose initial
preparation is under control and, after a given interaction time, the
excited state population is detected. The qubit is treated as a
normalized vector in a two-dimensional complex Hilbert space $\hil_Q$, with 
$\{\ket{e},\ket{g}\}$ providing an orthonormal basis. The qubit is
initially prepared in a pure state
\begin{equation}
\ket{\psi}=\cos \frac{\vartheta}{2} \ket{e}+e^{i\varphi}\sin \frac{\vartheta}{2}
\ket{g} \, ,
\end{equation}
with $\varphi \in \left[0,2\pi\right)$ and $\vartheta \in
\left[0,\pi\right]$. Hence the qubit density operator reduces to the
projector $\varrho_\Q=\ketbra{\psi}{\psi}$. Being a two-level system, by
appropriately choosing the zero energy level and denoting by $\omega$ its
transition frequency, the qubit Hamiltonian can be written as
\begin{equation*}
H_q=\frac{\hbar \omega}{2} \sigma_z \; .
\end{equation*}
The qubit-resonator interaction is the interaction between a
single-mode bosonic field and a two-level system. 
In the rotating-wave approximations and for the near-resonant 
case, i.e., for small values of the detuning $\delta=\omega - \Omega$
we have the Jaynes-Cummings (JC) model with Hamiltonian 
\begin{eqnarray}\label{JC}
\tilde{H}_{\JC}&=&H_q + H_\F + H_{int} \nonumber \\
&=&\frac{\hbar \omega}{2} \sigma_z + \hbar \Omega a^{\dagger}a 
+ \hbar \lambda\left(\sigma_+a + \sigma_-{a^{\dagger}}\right) \; .
\end{eqnarray}
The unperturbed Hamiltonian $\tilde{H}_{\JC}^{(0)}=H_q + H_\F $ satisfies
the eigenvalues equations
$$\tilde{H}_{\JC}^{(0)}\ket{k,n}=\hbar\left[n\Omega + \frac12\, \omega\,
(-1)^k \right]\ket{k,n}\,,$$ with $k=e,g$ and with 
the correspondences $0 \leftrightarrow e$, $1 \leftrightarrow g$.
In Eq. (\ref{JC}) $\lambda \in \R$ represents the coupling strength, 
$\sigma_+ a$ and $\sigma_-{a^{\dagger}}$ stand respectively for the 
operators $\sigma_+\otimes a$, $\sigma_-\otimes {a^{\dagger}}$ acting 
on the tensor product space, where $\sigma_{\pm}$ are the qubit 
ladder operators. Upon choosing a suitable rotating frame one rewrites
the Hamiltonian in interaction picture $H_{\JC}$: 
\begin{equation}\label{JCint}
H_{\JC}=\frac{\hbar \delta \sigma_z}{2} + 
\hbar \lambda\left(\sigma_+a + \sigma_-{a^{\dagger}}\right) \; .
\end{equation} 
The interaction only couples, for a given $n$, the states $\ket{e,n}$
and $\ket{g,n+1}$, and thus it is possible to study the interaction
inside the two-dimensional manifold spanned by these states leading to
a representation -- the so called dressed states basis -- where
$H_{\JC}$ is diagonal. We further assume the absence of any initial
correlations between the qubit and the oscillator, thus choosing at
time $t=0$ the following factorized density operator
$$
\varrho(0)=\varrho_\Q \otimes \varrho_{\F} \; ,
$$
whose dynamical evolution with respect to the JC Hamiltonian is
given by:
$$
\varrho(t)=U(t) \varrho(0){U}^{\dagger}(t) \;,
$$
with $U(t)=\exp{\left(-\frac{i}{\hbar} H_{\JC}t\right)}$.
\par
Time evolution entangles the qubit and the resonator \cite{sch10} and the
probabilities for the qubit to be found in the ground or excited state
are obtained via the Born rule as 
\begin{equation}\label{prob1}
p(j|\beta)=\Tr{\Q\!\F}{ \varrho(t)\ketbra{j}{j}\otimes {\mathbb I}_\F} 
\qquad j=e,g
\end{equation}
where $p(j|\beta)$ denotes the conditional probability of obtaining the 
value $j$ when the value of the temperature parameter is $\beta$. 
Upon introducing the following quantum operation:
\begin{equation}\label{probe1}
\varrho_\Q \stackrel{\mathcal{E}}{\longmapsto} 
\varrho_\p \equiv \Tr{\F}{U(t)\,\varrho_\Q\otimes
\varrho_\F\,{U}^{\dagger}(t)} \;,
\end{equation}
where $\mathcal{E} : \mathcal{L}(\hil_\Q) \rightarrow \mathcal{L}(\hil_\Q)$,
Eq.~(\ref{prob1}) can be equally rewritten at the level of the qubit 
subsystem alone, namely:
\begin{equation}\label{prob2}
p(j|\beta)=\Tr{\Q}{\varrho_{\p}\ketbra{j}{j}} \;.
\end{equation}
In the following we will refer to $\varrho_\p$ as the \textit{probe
state}: It describes the qubit subsystem at time $t$, obtained as 
the partial trace over the phonon field of the overall evolved state
of the coupled system. Since it is a density operator on 
$\hil_{Q}$ it can be arranged in a 2$\times$2 density matrix. We have 
\begin{eqnarray}
\varrho_\p&=&\sum_{n=0}^{\infty} p_n(\Omega,\beta)
\left(
\begin{array}{cc}
\varrho_{ee} & \varrho_{eg} \\ \varrho_{ge} & \varrho_{gg}
\end{array}
\right) \;, \nonumber 
\end{eqnarray}
where:
\begin{subequations}\label{matrixelement} \label{rhs} 
\begin{align}
 \ \varrho_{ee}&= \cos^{2}\frac{\vartheta}{2}\left[\cos^2\theta_n t +
 4\, \frac{\delta^2}{\theta_n^2}\sin^2\theta_n t\right]   \nonumber 
 \\ &\hspace{0.5cm}+ \sin^{2}\frac{\vartheta}{2}
 \frac{\lambda^2n}{\theta_{n-1}^2} 
 \sin^2\theta_{n-1}t,\\                     
\varrho_{eg}&= \frac12 e^{-i\varphi}\sin \vartheta
\left[ \cos\theta_{n-1}t +i\frac{2\delta}{\theta_{n-1}}\sin\theta_{n-1}t  
\right] \nonumber \\
 &\hspace{0.5cm}\times \left[ \cos\theta_n t -i\frac{2\delta}{\theta_n}
\sin\theta_n t \right],  \\
\ \varrho_{ge}&= \varrho_{eg}^{\ast} \quad \hbox{and} \quad
\ \varrho_{gg}= 1- \varrho_{ee},
\end{align}
\end{subequations} 
with: $$\theta_n\equiv \theta_n (\delta,\lambda)
=\frac12 \sqrt{\delta^2 +4\lambda^2\left( n+1\right)}\,.$$ 
\subsection{Effects of decoherence}
A purely Hamiltonian dynamics doesn't match realistic features. In
real-life scenarios quantum coherence is hard to achieve in
mechanical objects, and can be maintained for relatively small times
($\approx 10^{-9} $s ). Complete Rabi oscillations between the phonon
and the qubit excitation involve only the first Rabi half periods, then
a damping of the probabilities $p(j|\beta)$ to $\frac{1}{2}$ is
observed: the most striking signature of decoherence.  Hence we include
in our model the treatment of non dissipative decoherence occurring
between the qubit and the resonator. Following Ref. \cite{atomtrap} 
we consider an effective model provided by adding a power-law term in the thermal 
distribution, which leads to probe state matrix elements given by: 
$$ \tilde\varrho_{ij}=\sum_{n=0}^{\infty}p_n (\Omega,\beta) 
\left[ e^{-\gamma_nt}
\varrho_{ij} + \frac12 \left(1-e^{-\gamma_nt}\right)\right]
$$
being $\varrho_{ij}$ the matrix elements of Eq. (\ref{rhs}), as
evaluated for the unitary case,
$i,j \in \{e,g\}$ and $$\gamma_n=b(1+n)^a\,.$$ More explicitly 
\begin{subequations}\label{rhsd}
\begin{align}
\tilde\varrho_{ee}&=\frac{1}{2}
\left[1+\sum_{n=0}^{\infty} p_n(\Omega,\beta) 
e^{-\beta e^{-\gamma_nt}}
\left(\varrho_{ee}-\varrho_{gg}\right)\right], \\
\tilde\varrho_{eg}&= \frac{1}{2}\sum_{n=0}^{\infty}
p_n (\Omega,\beta)e^{-\gamma_nt}\varrho_{eg},\\
  \tilde\varrho_{ge}&=\tilde\varrho_{eg}^{\ast}\quad \hbox{and} \quad
  \tilde\varrho_{gg}= 1- \tilde\varrho_{ee}\,. 
\end{align}
\end{subequations}
One can see that the dynamical evolution now drives the qubit towards 
the maximally mixed state, described by the density operator $\frac{\mathbb I}{2}$.  
\section{Quantum thermometry}\label{s:QET}
In this section we apply the tools of (local) quantum estimation theory 
(QET) to the coupled qubit-oscillator system. An 
estimation problem always consists in two steps: at first one has to choose 
a measurement and then, after collecting a sample of outcomes, one
should find an estimator, i.e. a  function to process data and to infer 
the value of the quantity of interest.
In our case, temperature, expressed as 
$\beta$, is the unknown parameter which has to be estimated from the sample 
of outcomes coming from measurements performed on the qubit. The results, a 
string of zeroes and ones for the case of population measurement, are 
distributed according to the probabilities $p(j|\beta)\equiv\varrho_{jj}$ of 
Eqs. (\ref{prob2}) and (\ref{rhs}) [or Eq. (\ref{rhsd}) in  presence
of decoherence].  
The Cram\'er-Rao inequality establishes that the variance
Var$(\beta)$ of any unbiased estimator is lower bounded by
\begin{equation}\label{cramer-rao}
\mbox{Var}(\beta) \geq \frac{1}{M F(\beta)} \ ,
\end{equation}
where $M$ is the  cardinality of the sample, i.e., the number of 
measurements, and $F(\beta)$ the so-called Fisher information (FI):
\begin{eqnarray}\label{Fisher}
F(\beta)&=& \sum_{j=e,g} p(j|\beta)\left[\partial_{\beta}\ln
p(j|\beta)\right]^2 \nonumber \\ &=& \frac{{\left[ \partial_{\beta}
p(e|\beta)\right]}^2}{p(e|\beta)} +  \frac{{\left[ \partial_{\beta}
p(g|\beta)\right]}^2}{p(g|\beta)} \; .  
\end{eqnarray}
Efficient estimators are those saturating the Cram\'er-Rao inequality and
their existence depends on the statistical model. However, independently
of the statistical model we have that for sufficiently large samples, i.e., 
in the asymptotic regime $M\gg 1$, maximum likelihood estimators are
always efficient.
\par
Quantum mechanically, the probability of obtaining the outcome
$j\in\{e,g\}$ from a measurement is given according to the Born rule
by  $p(j|\beta)=\tr{\varrho_\p \Pi_j}$, where the probe state
$\varrho_\p\equiv \varrho_\p (\beta)$ parametrized by the unknown 
quantity $\beta$ is referred to as the quantum statistical model, and 
the collection of operators $\{\Pi_j\}$, $\Pi_j\geq 0$, $\sum_j\Pi_j=
\mathbb{I}$ is the probability operator-valued measure describing the 
measurement taking place on the qubit. In our
case the qubit excited state population is probed and the measurement
reduces to a projective one, $\ketbra{e}{e}$ and
$\ketbra{g}{g}=\mathbb{I}-\ketbra{e}{e}$, i.e., we are measuring 
the Pauli operator $\sigma_z=\ketbra{e}{e}-\ketbra{g}{g}$. 
\par
Once the observable is fixed, we optimize the estimation
procedure by maximizing the FI over 
the qubit state parameters, $\vartheta$ and $\varphi$, 
as well as over the parameters driving the interaction -- i.e., the detuning
$\delta$ and the interaction time $t$. In other words, by 
employing the optimal qubit preparation and tuning the interaction
parameters one may find a working regime achieving the maximum 
precision for that kind of measurement.  
\par
On the other hand, one may also maximize the FI over all possible
quantum measurements. Upon defining the symmetric logarithmic 
derivative (SLD) $L_{\beta}$ as the selfadjoint operator satisfying 
the equation
\begin{equation}
\frac{L_{\beta}\varrho_\p +\varrho_\p L_{\beta}}{2}=
\partial_{\beta}\varrho_\p \; ,
\end{equation}
it is possible to show that the Fisher information $F(\beta)$ of any quantum
measurement is upper bounded by the following quantity:
\begin{equation}\label{Ineq}
F(\beta)\leq G(\beta)\equiv \tr{\varrho_\p L_{\beta}^2} \; ,
\end{equation}
which is called quantum Fisher information (QFI). QFI does not depend on
the measurement carried on the qubit---indeed being obtained by maximizing
over the possible measurement. It is rather an attribute of the family
of states $\varrho_\p(\beta)$ parametrized by the temperature. Looking
back to the Cram\'er-Rao inequality Eq.(\ref{cramer-rao}) one sees that
QFI allows one to write its natural quantum version 
\begin{align}
\mbox{Var}(\beta) \geq \frac1{M G(\beta)}\,.\label{QCR}
\end{align}
The above equation represents the  Quantum Cram\'er-Rao bound (QCR),
i.e. the ultimate bound to the precision allowed by quantum mechanics 
for a given statistical model $\varrho_\p(\beta)$.  
An optimal measurement, i.e. a measurement whose FI $F(\beta)=G(\beta)$
equals the QFI for the parameter $\beta$, is given by the observable
corresponding to the spectral measure of the SLD $L_\beta$. On the other 
hand, other kind of measurements may achieve optimality for the whole range
of values of $\beta$ or for a subset of values. Indeed, we will see in
the following that population measurement is optimal for a suitable
choice of the initial qubit preparation. We remind that for the estimation 
of a single parameter, as it is in our case, the QCR  may be always attained, 
and an estimator saturating Ineq. (\ref{QCR}) is called efficient.
The existence of an efficient estimator depends on the statistical 
model. However, independently of the statistical model, for sufficiently 
large samples, i.e., in the asymptotic regime $M\gg 1$, maximum likelihood 
and Bayesian estimators are always efficient.
\par
Upon diagonalizing the probe state one achieves the decomposition 
$\varrho_\p=\varrho_+\ketbra{\psi_+}{\psi_+}\, 
+\, \varrho_-\ketbra{\psi_-}{\psi_-} $ and is able to solve the equation for SLD
\begin{align}
L_{\beta}=&\,\frac{\bra{\psi_+}\partial_{\beta}
\varrho_\p\ket{\psi_+}}{\varrho_+}\ketbra{\psi_+}{\psi_+}
 \nonumber \\ & +\,
\frac{\bra{\psi_-}\partial_{\beta}\varrho_\p
\ket{\psi_-}}{\varrho_-}\ketbra{\psi_-}{\psi_-}
 \nonumber \\ & +\, \frac{2}{\varrho_+ +
\varrho_-}\left[\bra{\psi_+}\partial_{\beta}
\varrho_\p\ket{\psi_-}\ketbra{\psi_+}{\psi_-}
 \right. \nonumber \\  & +\, \left.
\bra{\psi_-}\partial_{\beta}\varrho_\p
\ket{\psi_+}\ketbra{\psi_-}{\psi_+}
\right], \end{align}  
finally obtaining an explicit formula for the QFI
\begin{align}\label{qfi}
  G(\beta)=&\,
  \frac{\left(\partial_{\beta}\varrho_{+}\right)^2}{\varrho_{+}} +
  \frac{\left(\partial_{\beta}\varrho_{-}\right)^2}{\varrho_{-}}
  \nonumber \\
  &+\, 2\kappa\,\left[ \left| \langle
  \psi_-|\partial_{\beta}\psi_{+}\rangle \right|^2 + \left| \langle
  \psi_+|\partial_{\beta}\psi_{-}\rangle \right|^2 \right]
  \end{align}
where 
$$|\partial_{\beta}\psi_{\pm}\rangle = 
\partial_{\beta}\langle e|\psi_{\pm}\rangle\,|e\rangle 
+ \partial_{\beta}\langle g|\psi_{\pm}\rangle\,|g\rangle\,,$$ 
and 
$$\kappa=\frac{\left(\varrho_+ - \varrho_-
\right)^2}{\varrho_+ + \varrho_-}=(1-2\varrho_+)^2\,.$$ 
Eq. (\ref{qfi}) contains a first term which
resembles the FI and a second one, truly quantum in nature, which leads
to the QCR and vanishes whenever $\ket{\psi_{\pm}}$ does not depend on
$\beta$.
\section{Dynamics of the Fisher information and optimal 
working regimes}\label{s:results}
In this section we report results for the qubit-resonator coupled 
system with physical parameters chosen in a range matching the 
experimental setup of Ref. \cite{nmr11}. More specifically, we 
present a systematic study of the FI for population measurement as a
function of the state and interaction parameters, carrying out
numerical maximization and finding the optimal working
regimes. We also evaluate the QFI of the family of states
$\varrho_{P}(\beta)$ and find the ultimate bound to
precision, i.e. a benchmark 
in order to assess the performances of qubit thermometry via population measurement.
\par
Hereafter we work with dimensionless quantities by rescaling times and 
frequencies in units of the coupling $\lambda$.  We thus substitute
time, detuning and decoherence parameters by their rescaled counterparts
\begin{align}
t \longmapsto \tau \equiv \lambda t \notag, \quad
\delta \longmapsto \gamma \equiv \delta/\lambda \notag, \quad
b \longmapsto \tilde{b}\equiv b/\lambda\,. \notag
\end{align} 
Effective detuning $\gamma$ will range in $|\gamma| \in [0, 1.5]$. Also
a dimensionless effective temperature $\tilde{\beta}$ is defined,
provided by the substitution 
$$\beta \longmapsto \tilde{\beta}\equiv\beta\hbar\Omega\,.$$ 
For convenience, we continue to term
$\tilde{\beta}$ and $\tilde{b}$ respectively $\beta$ and $b$.
\subsection{Resonant Hamiltonian regime}
Upon using the expression of the diagonal matrix elements in Eqs.
(\ref{rhs}) we have evaluated the FI of Eq. (\ref{Fisher}).  
We start the discussion by considering the resonant case, i.e 
zero detuning, and analyze the effect of detuning afterward in this 
Section.  For convenience we
adopt the notation $F(\beta)$ for the FI, but keep in mind the
complete dependence $F(\beta;\vartheta,\tau,\gamma)$ on both the qubit
degrees of freedom and the parameters $\gamma$ and $\tau$ which drive
the coupling.  Notice that $F(\beta)$ does not depend on the qubit phase $\varphi$: its
building-blocks are in fact the probabilities $p(e|\beta)$ and
$p(g|\beta)$, whereas $\varphi$ only appears in off-diagonal matrix
elements. Varying the parameter $\vartheta$ from $\pi$ to $0$ we span
the entire class of qubit preparation, starting from $\ket{1}$, going
trough a superposition and ending in $\ket{0}$.
\begin{figure}[h!] 
\includegraphics[width=0.85\columnwidth]{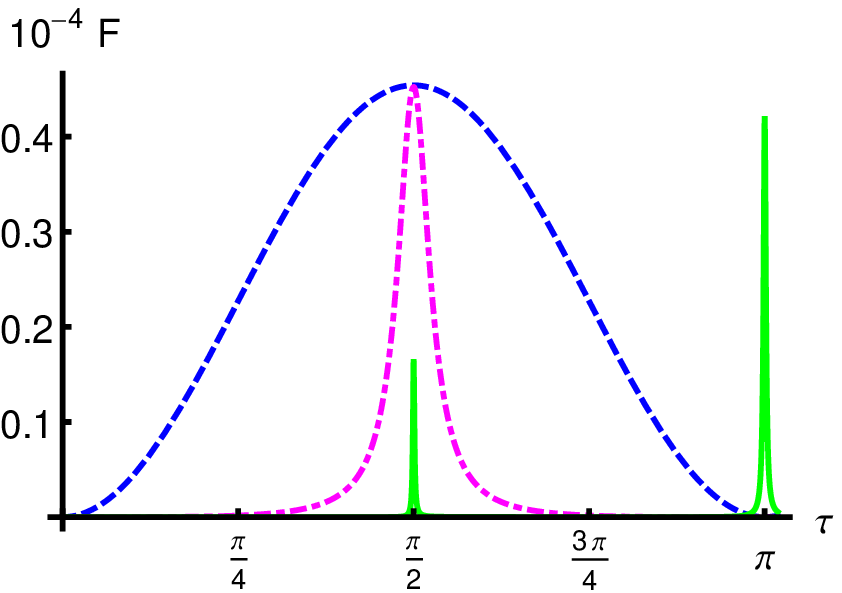} 
\includegraphics[width=0.85\columnwidth]{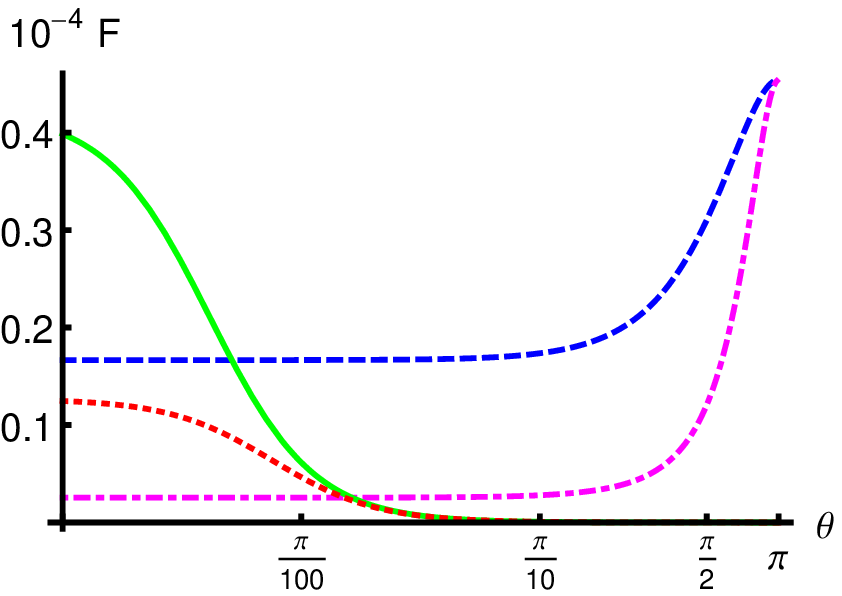} 
\caption{(Color online) Upper panel: FI for $\beta=10$ as a function of
the effective time $\tau$, for different $\vartheta$ values:
$\vartheta=\pi$ (dashed blue), $\vartheta=0.95\, \pi$ (dot-dashed magenta) and
$\vartheta=0$ (solid green). FI takes a pronounced global maximum at
$(\vartheta,\tau)=\left(\pi,\frac{\pi}{2}\right)$ while it
is possible to see a secondary extremely peaked maximum, which
occurs for $\tau=\pi$ and preparing the qubit in $\ket{0}$. Lower
panel: log-linear plot of the FI for $\beta=10$ as a function of $\vartheta$ 
for, $\tau=\frac{\pi}{2}$ (dashed blue),
$\tau=\frac{\pi}{2}+\varepsilon$ (dot-dashed magenta),
$\tau=\pi$ (solid green),
$\tau=\pi+\varepsilon$ (dotted red), with $\varepsilon = 0.01$.
\label{f:fish1}}
\end{figure}
\par
Let us now consider the system at a fixed value of the temperature,
e.g. where the resonator is supposed to be very close to the
ground state, say $\beta=10$. The probabilities
$p(j|\beta)=\varrho_{jj}$ evolve periodically in time
according to Eq. (\ref{rhs}), as the coupled system undergoes Rabi
oscillations. The corresponding behavior of the
FI is shown in the upper panel of Fig.~\ref{f:fish1}. The FI displays 
a robust maximum at the optimal time $\tau_{\rm max}=\frac{\pi}{2}$ for 
$\vartheta=\pi$, corresponding to prepare the qubit in its ground state. 
This maximum is, at the same time, the global and the smoothest one. In
fact, as soon as $\vartheta$ is moved from $\pi$ the FI
suddenly drops to zero, except for a sharp peak centered in
$\tau_{\rm max}$, monotonically decreasing with respect to $\vartheta$, as
shown in the lower panel of Fig.~\ref{f:fish1}.  Another maximum of 
the same order of the global one can be found at $(\vartheta,\tau)=
\left(0, \pi \right)$  but 
it is extremely peaked, thus representing a bad (unstable) choice for 
a possible measurement. Upon inspecting the temporal evolution of the 
excited state probability we found that $p(e|\beta)$ has a minimum at
$\tau=\tau_{\rm max}$, a fact which gives us a physical insight on the FI
behavior: since our goal is the estimation of a vanishing quantity which
carries information about thermal disorder, we expect to find the maximum
sensitivity in our predictions where the excitation is most likely
stored -- as a phonon -- in the resonator, i.e., when $p(e|\beta)$ is
minimum.
\begin{figure}[h!] 
\includegraphics[width=0.85\columnwidth]{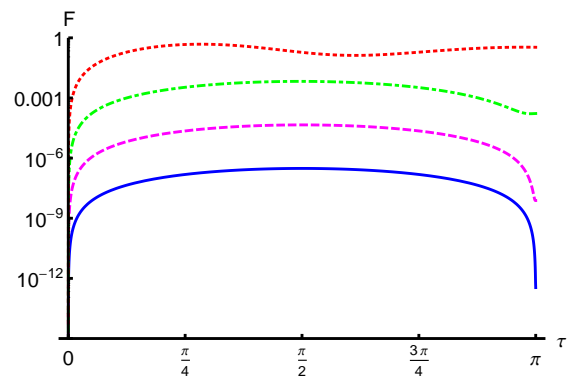} 
\caption{(Color online) Log-linear plot of the FI as a function of
  effective time $\tau$ for different values of $\beta$. The qubit 
  is prepared in the ground state $\ket{1}$ ($\vartheta=\pi$). From bottom to top
  $\beta=15$ (solid blue), $\beta=10$ (dashed magenta), $\beta=5$ (dot-dashed green),
  $\beta=1$ (dotted red). Upon raising the temperature the FI no longer keeps
  a scale-free shape: thermal excitations modifies its profile making
  it irregular. In particular the global maximum comes earlier in
  time.
\label{f:fish3}} \end{figure}
\par
Let us now turn our attention to the dependence of the FI
on the temperature itself. In Fig.~\ref{f:fish3} we show, on a logarithmic 
scale, the temporal evolution of the FI for different values of $\beta$. 
FI varies over several orders of magnitude, matching our intuition 
that the closer we are to the ground state, the harder is to achieve a
given precision in estimation of temperature. Furthermore, upon 
lowering the temperature, the temporal evolution of $p(j|\beta)$ becomes
less involved, finally approaching the exactly periodic one of Rabi 
oscillations, which in turn freezes the profile of the FI in a 
shape independent on the temperature itself.
\par
\begin{figure}[h!] 
\centering 
\includegraphics[width=0.91\columnwidth]{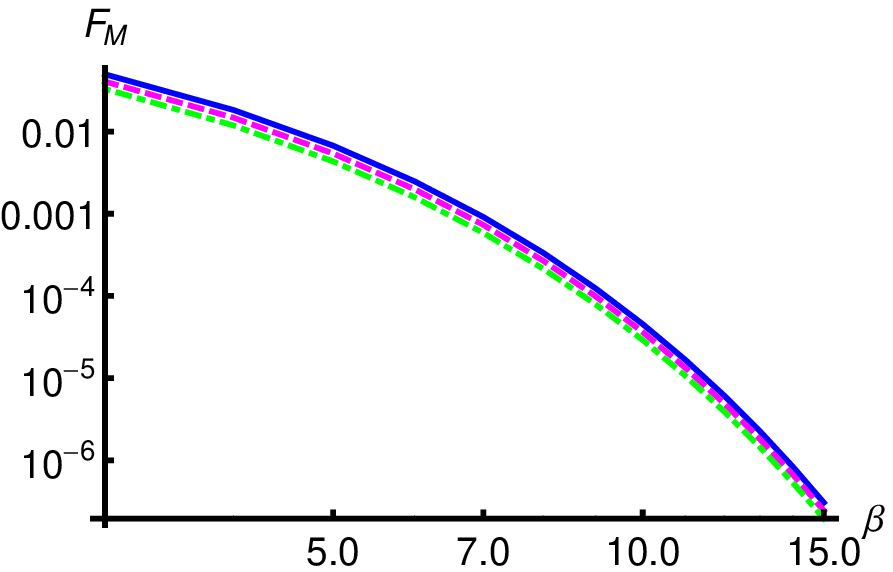}
\includegraphics[width=0.854\columnwidth]{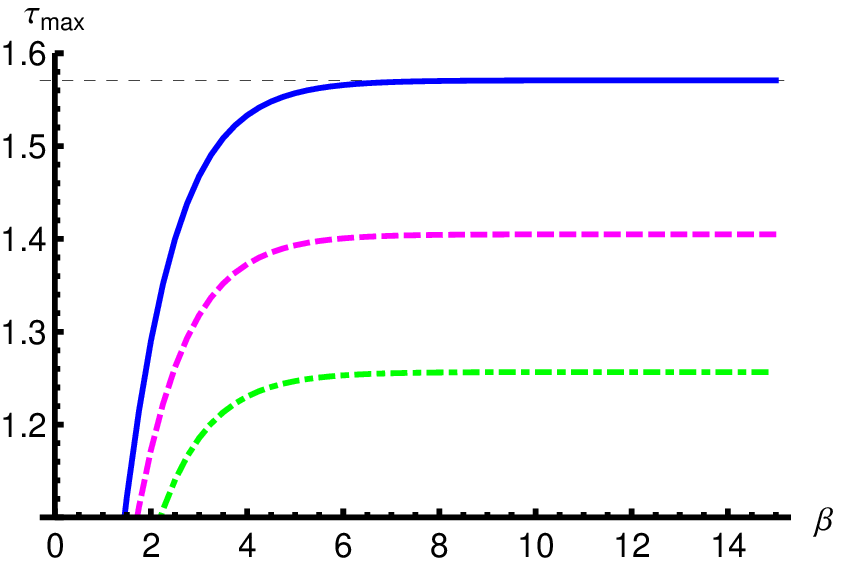} 
\caption{(Color online) Upper panel: log-log plot of the FI maximized
over   $\tau$ as a function of $\beta$, with $\theta=\pi$ for different
values of   detuning: $\gamma=0$ (solid blue), $\gamma=1$ (dashed
magenta), $\gamma=1.5$   (dot-dashed green). Bottom panel: the times
$\tau_{\rm max}$ which maximizes   the FI as a function of $\beta$, with
$\theta=\pi$ for different values of  $\gamma$ (same values and colors
of the upper panel).\label{f:max}} 
\end{figure}

The qubit preparation $\theta=\pi$ is universally optimal, i.e., it
leads to a maximum of the FI independently of the interaction time.
After fixing $\theta=\pi$ we have numerically maximized $F(\beta)$
with respect to $\tau$. The solid blue line of the upper panel of 
Fig.~\ref{f:max} is the the log-plot of 
$$F_M (\beta) = \max_\tau F(\beta)\,,$$ 
as a 
function of $\beta$, from which it is apparent the exponential decrease 
of the maximum value
achieved by the FI for increasing $\beta$. The Cram\'er-Rao inequality
immediately relates this fact to an exponential loss of sensitivity
moving towards the quantum ground state of the resonator. An other
interesting feature that emerges from the maximization is a shift in
the value of the optimal interaction time. In the lower panel of
Fig.~\ref{f:max} we can recognize the existence of a steady value for
the optimal time $\tau_{\rm max}=\frac{\pi}{2}$ when approaching the
ground state, while for smaller values of $\beta$ the optimal time
comes earlier. In fact, the temporal evolution of FI (see Fig.
\ref{f:fish3}) not only predicts
an exponential increase of the global maximum when temperatures are
raised, but also a shift of its location.
\par 
\subsection{Effects of detuning}
In this section we take into account the possible existence 
of a nonzero detuning $\gamma$ between the oscillator and the 
qubit frequencies. This has two main consequences, which are both 
illustrated in Fig.~\ref{f:max}. On the one hand, the maximum achievable 
value of the FI slightly decreases and, on the other hand, the optimal
interaction time $\tau_{\rm max}$ at which the maximum takes
place anticipates. Therefore, the best working conditions to achieve 
the optimal sensitivity in the estimation of $\beta$ correspond to 
have the qubit and the resonator in resonance.  It is also
worth to notice  that $\gamma$ does not represent a critical parameter, 
as the initial preparation of the qubit, 
since the FI dependence on $\gamma$ is smooth. One can see this in the
upper panel of Fig.~\ref{f:max}, where we see that curves corresponding
to quite different values of the detuning are almost superposed.
\subsection{Quantum Fisher information}
In order to assess the performances of the population measurement
in the estimation of temperature we have evaluated the QFI of the family
$\varrho_\p(\beta)$. The diagonalization of the probe state has to be
carried out numerically, hence in general analytical expressions of the
QFI are not available. A first fact is that $G(\beta)$ turns
out to be independent on the qubit phase $\varphi$, which then does not
represent an extra degree of freedom whereby gain more restrictive
bounds to precision on $\mbox{Var}(\beta)$. Even the optimal qubit
preparation for to the best conceivable measurement involves
control of the parameter $\vartheta$ only. 
\par
\begin{figure}[h!] 
\centering 
\includegraphics[width=0.85\columnwidth]{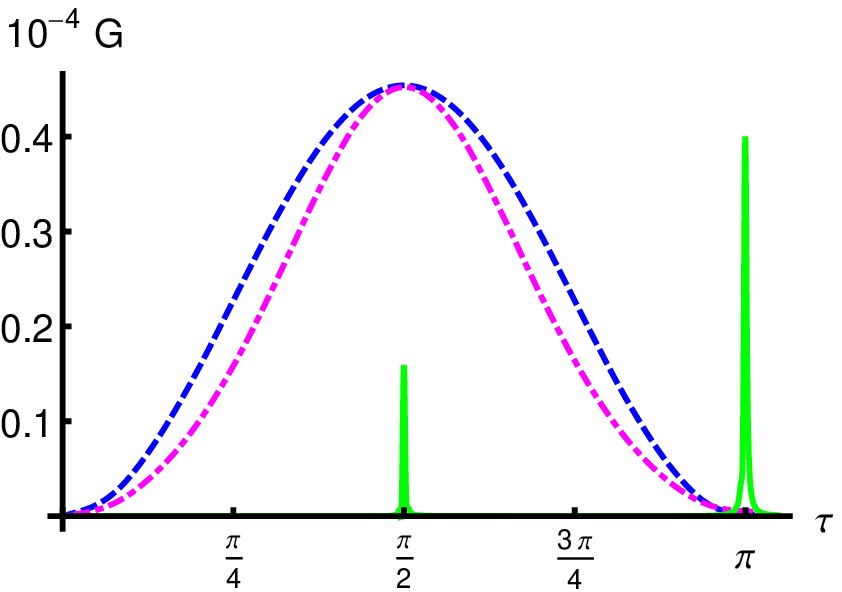}
\includegraphics[width=0.85\columnwidth]{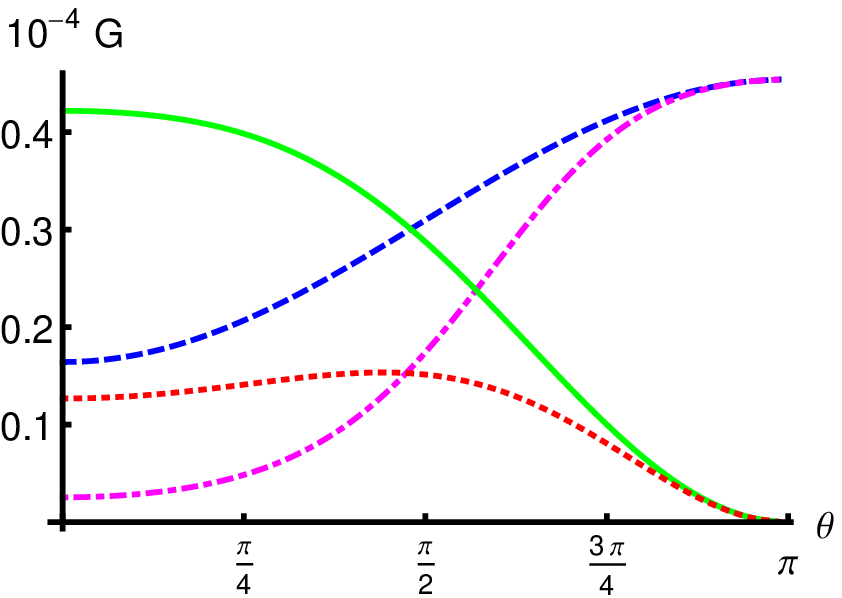} 
\caption{(Color online) Upper panel : QFI for
  $\beta=10$ as a function of $\tau$, for $\vartheta=\pi$ (dashed blue),
  $\vartheta=0.95\, \pi$ (dot-dashed magenta) and $\vartheta=0$ (solid green). QFI
  behaves like FI for $\vartheta=\pi$ leading to the same maximum,
  while for smaller angles it shows a smoother profile. For angles
  $0<\vartheta<\pi$ one may find measurements which improve the
  precision of temperature estimation. 
  Bottom
panel: QFI for $\beta=10$ as a function of $\vartheta$ 
for
$\tau=\frac{\pi}{2}$ (dashed blue),
$\tau=\frac{\pi}{2}+\varepsilon$ (dot-dashed magenta),
$\tau=\pi$ (solid green),
$\tau=\pi+\varepsilon$ (dotted red), with $\varepsilon = 0.01$. \label{f:qfi}}
\end{figure}

As we have done for the FI, we start to inspect the QFI behavior for a fixed
value of temperature $\beta$ in the resonant case. Also for the QFI
the maximum is achieved by preparing the qubit in the state $\ket{g}$
and probing it at time $\tau_{\rm max}$. In this case the behavior of
$G(\beta)$ is identical to that of $F(\beta)$, as it is apparent by
comparing Figs. \ref{f:fish1} and \ref{f:qfi}. In other words, for a
given value of the parameter $\beta$ into the range explored, the
choice $(\vartheta,\tau)=(\pi,\tau_{\rm max})$ makes population
measurement optimal. Moreover, the QFI itself reaches its global
maximum for that choice. Thus, provided that an optimal estimator is
employed, e.g. maximum likelihood in the asymptotic regime, this
strategy provides optimality in sense that either inequality
(\ref{Ineq}) is saturated and the right-hand side of QCR is as low
as possible.
\par
This conclusion is confirmed upon a closer inspection of the probe state. When
$\vartheta=\pi$ the off-diagonal terms vanish and $\varrho_\p$ is
diagonal, with eigenvalues 
\begin{subequations}
\begin{align}
\varrho_+&=  \sum_{n=0}^{\infty} p_n(\Omega,\beta) 
\sin^2\left[\sqrt{\gamma^2+4n}\,\frac{\tau}{2}\right] \frac{n}{n+\gamma^2/4} \\
\varrho_-&=  1-\varrho_+
\end{align}
\end{subequations}
As a consequence, the QFI reduces to 
$$G(\beta;\pi,\tau,\gamma)=\frac{\left(\partial_{\beta}
\varrho_{+}\right)^2}{\varrho_{+}} + 
\frac{\left(\partial_{\beta}\varrho_{-}\right)^2}{\varrho_{-}}\,,$$ 
which coincides with the FI ruling the estimation of
$\beta$ via population measurement. \par 
On the other hand, some striking difference emerges between the
performances of population measurement and that of the optimal one if
the qubit is not prepared in the optimal (ground) state.
\par
In the lower panel of Fig.~\ref{f:qfi} we show $G(\beta)$ as a function of $\tau$ for 
different values of $\vartheta$: for $\vartheta<\pi$ the decrease of
$G$ is definitely smoother than that of $F$ and thus, in principle, some
measurement may be found making the initial preparation a less critical
parameter. Moreover inspecting the cut of the QFI along
$\tau=\pi$ we note that the maximum in $\vartheta=0$ becomes more
achievable compared to the one of $F(\beta)$.  All these features
suggest that for qubit preparations different from the ground state 
there will
be a sensible difference between the precision provided by population 
measurement and the optimal one implementable on the system.
On the other hand, being the overall maximum achievable with population
measurement, our results indicate that the
achievement of the ultimate bound to precision allowed by quantum
mechanics is in the capabilities of the current technology. 
\subsection{Effects of decoherence}
In this section we discuss the solution of the reduced qubit 
dynamics in the presence of dissipative decoherence, see Eq. (\ref{rhsd}), 
and inspect the corresponding  behavior of the FI. For the sake of
simplicity we consider zero detuning. Analogue results are obtained when 
including the detuning.
\par
The probabilities $p(j|\beta)=\tilde\varrho_{jj}$ are 
damped so that, waiting for a sufficient long time, whose value
depends on $a$ and $b$, we would find them to be identically $1/2$ or,
equally stated, the dynamical evolution brings the state to the maximally
mixed one. The contribution of decoherence is of the kind
exp$\left[-b(1+n)^a\tau \right]$ for every $n$, where $b$ has been
rescaled in coupling units $b\longmapsto b/\lambda$. Being a
multiplicative coefficient, as soon as $b$ is different from zero, the
exponential term will participate in killing the sums. Our calculations
show a relevant dependence of the FI on the parameter $b$, namely
values $b\approx 10^{-5}$ are sufficient to produce visible effects,
while varying $a$ in the range $(0,1)$ does not
deeply influence of FI behavior.
\par
\begin{figure}[h!] 
\includegraphics[width=0.85\columnwidth]{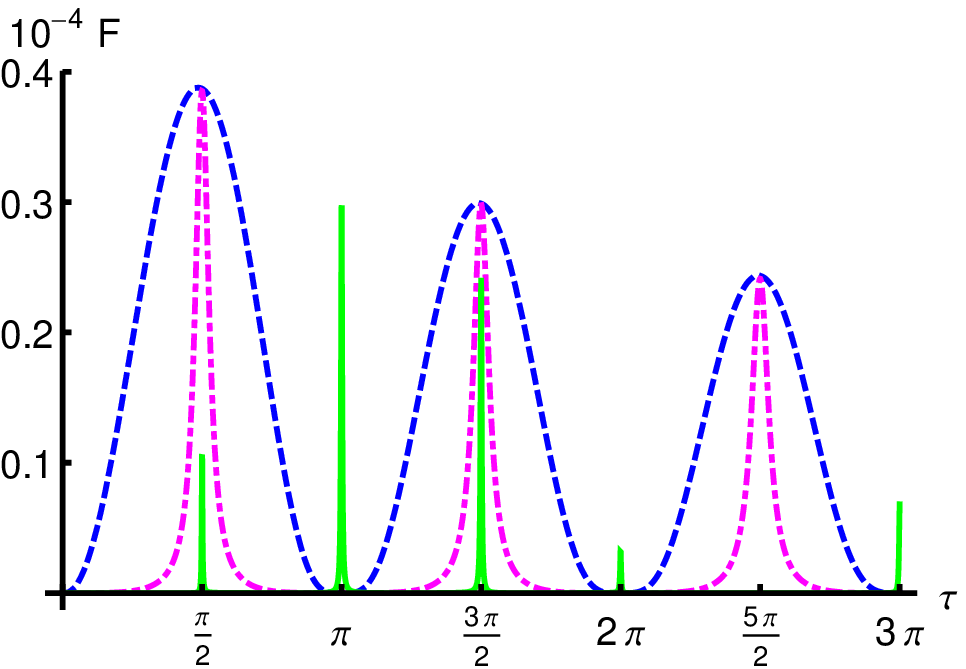} 
\includegraphics[width=0.9\columnwidth]{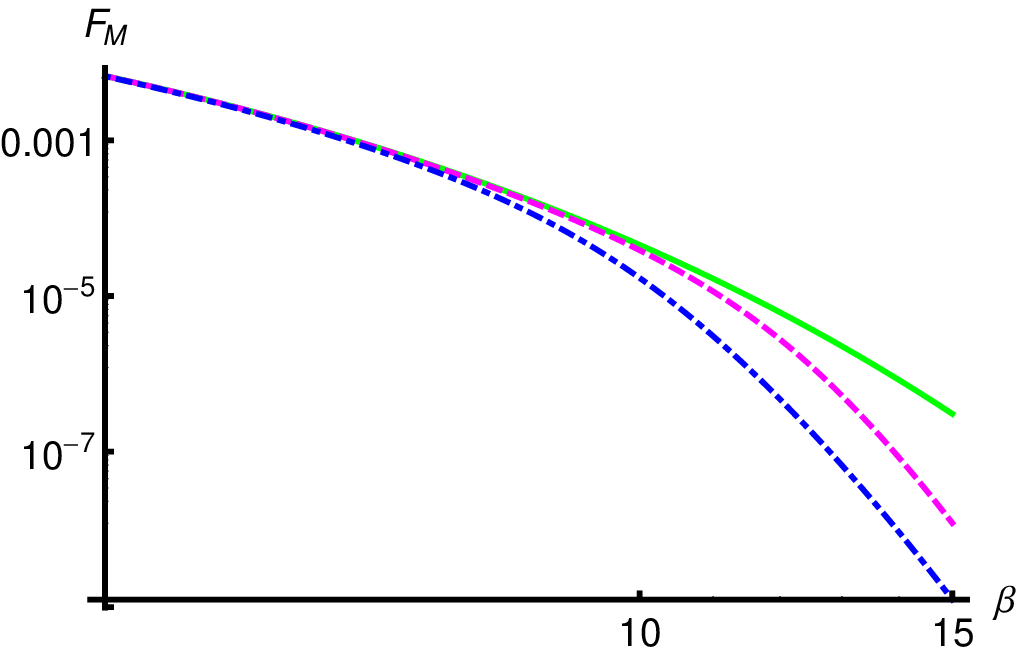} 
\caption{(Color online) Upper panel: Fisher information $F(\beta)$ for $\beta=10$ 
as a function of $\tau$ in the presence of decoherence and 
for different qubit preparations. The decoherence parameters 
are chosen as to $a=0.1$ and $b=10^{-5}$. Dashed blue line stands for $\vartheta=\pi$,
dot-dashed magenta for $\vartheta=0.95\,\pi$ while solid green ones for $\vartheta=0$. 
Having included decoherence treatment enables us not to restrict 
the evolution to the first Rabi half-period.
Lower panel: log-log plot of the Fisher information $F_M (\beta)$ maximized over the
interaction time, and in the presence of decoherence, as a function 
of $\beta$ and for fixed $\vartheta=\pi$,  for
 $b=0$ (solid green), $b=10^{-5}$ (dashed magenta), $b=10^{-4}$ (dot-dashed blue).\label{f:fishdec1}}
\end{figure}

In Fig.~\ref{f:fishdec1} we show the temporal evolution of the FI for
$\beta=10$, in the presence of decoherence and for different initial
preparations of the qubit. In the Hamiltonian regime
for large $\beta$ the resonator is close to the ground state, 
the evolution of $p(j|\beta)$ is periodic and hence, due to 
Eq. (\ref{Fisher}), the same is true for the FI. Upon incorporating 
decoherence we see that FI decays at a rate depending on $b$ and thus 
an irreversible dynamics emerges, which matches the physical evidence 
of a limited coherence time. On the other hand, a clear maximum at 
$\tau=\pi/2$ still appears, with a slightly decreased value of
$F(\beta)$. In the lower panel of  Fig.~\ref{f:fishdec1} we show the 
maximum value $F_M = \max_\tau F(\beta)$ for different values of the decoherence 
parameter. As it
is apparent from the (log-log) plot for high temperature (smaller $\beta$)
the effect of decoherence is negligible, whereas for increasing $\beta$
the effect is becoming more and more relevant.
\par
\section{Conclusions}
\label{s:out}
The temperature of a physical object cannot be directly measurable.  On
the other hand is can be regarded as a parameter whose value can be
indirectly inferred by measuring some proper observable and then
suitably processing the outcomes, an inference procedure usually
referred to as an estimation procedure. In the case of a micromechanical
oscillator with an isolated vibrational mode, effective schemes have
been suggested and realized \cite{nmr11} which rely on coupling the
resonator to a superconducting qubit and probing the latter using
population measurements. In other words, the qubit is employed as a
quantum thermometer to demonstrate that the resonator has been cooled to
its quantum ground state. In this paper we have analyzed in details
qubit thermometry in these systems, i.e., the estimation of temperature
via quantum limited measurements performed on the qubit.  In the
framework of quantum estimation theory we have analyzed precision as a
function of both the qubit initial preparation and the interaction
parameters, and we have evaluated the limits to precision posed by
quantum mechanics to qubit thermometry. 
\par 
We have computed the FI for population measurement, which is the 
appropriate figure of merit to assess the precision of estimation, 
and have found that its maximum, and hence the minimum variance in 
the estimated temperature, is achieved by preparing the qubit in 
the ground state, and probing it at an emergent time $\tau_{\rm max}$, 
which is predictable.  Furthermore, we have analyzed in details how 
the maximum depends on the temperature itself, on the detuning, and on the
noise parameter when one takes into account non dissipative decoherence.
In order to evaluate the ultimate bound allowed by quantum mechanics
to the sensitivity of temperature estimation, we have also computed the
quantum Fisher information. We found that QFI is maximized for the same choice of qubit
preparation and measurement time of the FI, and that for these common
values the maxima of FI and QFI coincide. We thus conclude that
population measurement is optimal for temperature estimation. 
\par
The range of parameters addressed in our analysis is that 
of recent experimental implementations \cite{nmr11}. We thus conclude 
that optimal estimation of temperature can be done with 
current technology. Since the FI of population measurement, 
and the QFI of the model, both decrease with the decrease of 
temperature, the estimation of lower temperature will be 
intrinsically less precise. On the other hand, since the are 
regimes, also in the presence of decoherence, where the maxima 
of the FI and the QFI are reasonably smooth as a function of 
the qubit preparation and of the interaction time we do 
not expect any "no-go" theorem for temperature estimation. 
In other words, we expect that optimal estimation of lower 
resonator temperatures, perhaps achievable with further experimental 
advances, will be still possible with population measurements. On the
other hand, ``optimality'' will correspond to an inherently less precise 
procedure compared to the case of higher temperature.
\par
Our analysis shows the optimality of feasible qubit thermometry in
providing quantum benchmarks for high precision temperature measurement,
as well as an efficient operational quantification of temperature for
mechanical modes lying arbitrary close to their ground state.  In other
words, achievement of the ultimate bound to precision allowed by quantum
mechanics is in the capabilities of the current technology.  Our results
also confirm that QET is a useful tool for assessing and comparing
inference procedures arising in quantum limited measurements \cite{sta10}, 
even when mesoscopic objects are involved.
\section*{Acknowledgments}
This work has been partially supported the CNR-CNISM agreement.

\end{document}